\documentclass[aps,prl,twocolumn,showpacs,superscriptaddress]{revtex4}
\usepackage{epsfig}

\newcommand{\ket}[1]{\ensuremath{|#1\rangle}}
\newcommand{\be}{\begin{equation}}
\newcommand{\ee}{\end{equation}}
\newcommand{\ba}{\begin{eqnarray}}
\newcommand{\ea}{\end{eqnarray}}
\graphicspath{{pics/}}
\def\lsim{\mathrel{\rlap{\lower4pt\hbox{\hskip0pt$\sim$}}
    \raise1pt\hbox{$<$}}}
\def\gsim{\mathrel{\rlap{\lower4pt\hbox{\hskip0pt$\sim$}}
    \raise1pt\hbox{$>$}}}
\DeclareMathSizes{10}{10}{5}{3}

\begin{document}

\title{Decoherence and disorder in quantum walks: From ballistic spread to localization}

\author{A. Schreiber}  
	\email{Andreas.Schreiber@mpl.mpg.de}
\affiliation{Max Planck Institute for the Science of Light, G\"{u}nther-Scharowsky-Str. 1 / Bau 24, 91058 Erlangen, Germany.}
\author{K. N. Cassemiro}
\affiliation{Max Planck Institute for the Science of Light, G\"{u}nther-Scharowsky-Str. 1 / Bau 24, 91058 Erlangen, Germany.}
\author{V. Poto\v{c}ek}
\affiliation{Department of Physics, FNSPE, Czech Technical University in Prague, B\v rehov\'a 7, 115 19 Praha, Czech Republic.}
\author{A. G\'abris}
\altaffiliation{Secondary address: Research Institute for Solid State Physics and Optics, Hungarian  Academy of Sciences, H-1525 Budapest, P. O. Box 49, Hungary.}
\affiliation{Department of Physics, FNSPE, Czech Technical University in Prague, B\v rehov\'a 7, 115 19 Praha, Czech Republic.}
\author{I. Jex}   
\affiliation{Department of Physics, FNSPE, Czech Technical University in Prague, B\v rehov\'a 7, 115 19 Praha, Czech Republic.}
\author{Ch. Silberhorn}
\affiliation{Max Planck Institute for the Science of Light, G\"{u}nther-Scharowsky-str. 1 / Bau 24, 91058 Erlangen, Germany.}
\affiliation{ University of Paderborn, Applied Physics, Warburger Str. 100, 33098 Paderborn, Germany.}
\date{\today}

\begin{abstract}
We investigate the impact of decoherence and static disorder on the dynamics of quantum particles moving in a periodic lattice. Our experiment relies on the photonic implementation of a one-dimensional quantum walk. The pure quantum evolution is characterized by a ballistic spread of a photon's wave packet along 28 steps. By applying controlled time-dependent operations we simulate three different environmental influences on the system, resulting in a fast ballistic spread, a diffusive classical walk and the first Anderson localization in a discrete quantum walk architecture.
\end{abstract}

\maketitle

Random walks describe the probabilistic evolution of a classical particle in a structured space resulting in a diffusive transport. In contrast, endowing the walker with quantum mechanical properties typically leads to a ballistic spread of the particle's wave function \cite{Aharonov}. The coherent nature of quantum walks has been theoretically explored, providing interesting results for a wide range of applications. They state not only a universal platform for quantum computing \cite{Childs} but also constitute a powerful tool for modelling biological systems \cite{Mohseni,Whaley, Aspuru-Guzik}, thus hinting towards the mechanism of energy transfer in photosynthesis. Quantum walks of single particles on a line have been experimentally realized in several systems, e.g. with trapped atoms \cite{Meschede} and ions \cite{Schaetz,Blatt}; energy levels in NMR schemes \cite{NMR,NMR2}; photons in  waveguide structures \cite{Silberberg2}, a beam splitter array \cite{White}, and in a fiber loop configuration \cite{Schreiber}.
Although these experiments opened up a new route to higher dimensional quantum systems, more sophisticated quantum walks need to be implemented to pursue the realm of real applications. A first step in this direction has been recently reported \cite{OBrien}, in which two particles execute a simultaneous walk and display intrinsic quantum correlations. 

One of the most important requirements for realizing quantum walk-based protocols is the ability to control the dynamics of the walk, that is to access and manipulate the walker's state in a position dependent way \cite{SKW,Ambainis}. In this paper we present the first experimental realization of quantum walks with tunable dynamics. We investigate the evolution of quantum particles moving in a discrete environment presenting static and dynamic disorders.

As predicted by Anderson in 1958 \cite{Anderson}, static disorder leads to an absence of diffusion and the wave function of the particle becomes localized, which, e.g. would render a conductor to behave as an insulator. Anderson localization has been experimentally investigated in different physical scenarios, e.g. employing photons moving in semiconductor powders \cite{Righini} and photonic lattices \cite{Schwartz, Silberberg}, or even via Bose-Einstein condensates \cite{Roati, Aspect}. However, although theoretically predicted in the context of quantum walks \cite{Toermae,Winter,Evangelou}, the effect has never been observed in a discrete quantum walk scenario.

Furthermore, it is interesting to note that the energy transport in photosynthetic light-harvesting systems is influenced by both, static and dynamic disorders, and it is precisely the interplay between the two effects that lead to the highly efficient transfer in those molecular complexes \cite{Whaley,Aspuru-Guzik}. Thus, in order to simulate a realistic influence of the environment, we go further in our studies by investigating the effect of dynamical noise, which typically induces decoherence \cite{Ambainis 2, Kendon}. Utilizing the ability to easily tune the conditions for the quantum walk, we demonstrate here the diverse dynamics of quantum particles propagating in these different systems.

In our experiment we realize the quantum walk of photons by employing a linear optical network. The evolution of the particle's wave function $\ket{\psi(x)}$ is given by  
\begin{equation}
\ket{\psi(x)}\rightarrow \gamma_x \ket{\psi(x)} + \sum_{k\neq x}\beta_{x,k}\ket{\psi(k)},
\label{evol}
\end{equation}
with the position dependent amplitudes $\gamma_x$ and $\beta_{x,k}$ determining the probability of the particle to stay at the discrete position $x$ or evolve to the adjacent sites $k$, respectively.

We study the expansion of the particle's wave packet in four different scenarios. (i) First of all we implement the quantum walk in a homogeneous lattice, showing that it presents an evolution that is free from decoherence. (ii) Next, we introduce static disorder by manipulating the lattice parameters $\gamma_x$ and $\beta_{x,k}$ , thus observing Anderson localization. We then examine two scenarios leading to decoherence, which essentially differ in the time scales of the occurring dynamic perturbations. (iii) In this case a dynamic randomization of the lattice parameters $\gamma_x$ and $\beta_{x,k}$ simulates the evolution of a particle interacting with a fast fluctuating environment. The resulting dephasing suppresses the underlying interference effects and hence causes the particle to evolve just like in a classical random walk \cite{Blatt,Meschede}. (iv) In the last scenario we simulate a slowly changing homogeneous environment. While  $\gamma_x$ and $\beta_{x,k}$ are stable during a single realization, a slow drift leads to different conditions for subsequent particles, thus affecting results obtained in an ensemble measurement.

In a discrete quantum walk the position of a particle evolves according to its internal coin state $\ket{c}$. For our photonic implementation we use the linear horizontal $\ket{H}=(1,0)^{T}$ and vertical $\ket{V}=(0,1)^{T}$ polarization of light. The state of the photon after N steps of the walk is found by applying the unitary transformation $U = \prod_{n=1}^N \hat{S}\hat{C_n}$ to the initial state $\ket{\psi(x)_{0}} = \ket{x_0}\otimes\ket{c_0}$. The coin operation $\hat{C}_n(x)$ manipulates the polarization of the photon in dependence on the position $x$ and the step number $n$. In the basis $\{\ket{H}, \ket{V}\}$ the coin operator is given in matrix form by 
\begin{equation}
 C(x) = \left(
                \begin{array}{cc}
                           e^{i\mathrm{\phi}_{\mathrm{H}}(x)} & 0  \\
                           0 & e^{i\phi_{\mathrm{V}}(x)} 
                 \end{array}
               \right)
		\left(\begin{array}{cc} \cos(2\theta) & \sin(2\theta) \\ \sin(2\theta) & -\cos(2\theta)
                 \end{array}\right),
\label{coin}
\end{equation}
with the diagonal matrix representing a phase shift $\phi_\mathrm{H}(x)$ for horizontal and $\phi_\mathrm{V}(x)$ for vertical polarizations, while the second matrix corresponds to a polarization rotation of $2\theta$.
The step operation $\hat{S}$ shifts the position $x$ of the photon by $+ 1$ if the polarization is horizontal and by $-1$ if it is vertical.

Following Eq.(\ref{evol}), the evolution of the wave function with the step number $n$ is given by
\begin{equation}
\ket{\psi(x)_{n+2}} = \gamma_x \ket{\psi(x)_n} + \beta_{x,x\pm2} (\ket{\psi(x+2)_n}+ \ket{\psi(x-2)_n}).
\label{schroedinger}
\end{equation}
Note that the transition coefficients $\gamma_x$ and $\beta_{x,x\pm2}$ are fully set by the coin operations $C_{n+1}(x)$ and $C_{n+2}(x)$. By changing the parameters $\phi(x)_\mathrm{H/V}$ and $\theta$ in a controlled way we can alter the coefficients and hence create diverse types of physical conditions for a quantum walk scenario.

A simple measure to quantify the spread of the wave function in the different systems is provided by the variance $\sigma^2$ of the final spatial distribution. While the decoherence free quantum walk presents a ballistic spread, with $\sigma^2 \propto n^2$, the classical random walk is diffusive, characterized by $\sigma^2 = n$. In contrast to both, in a one dimensional system with static disorder the wave packet shows exponential localization after a short initial expansion. The stagnation of the wave packet spread is thus evidenced by a constant variance.

\begin{figure}
\centering 
\includegraphics[width=0.5\textwidth]{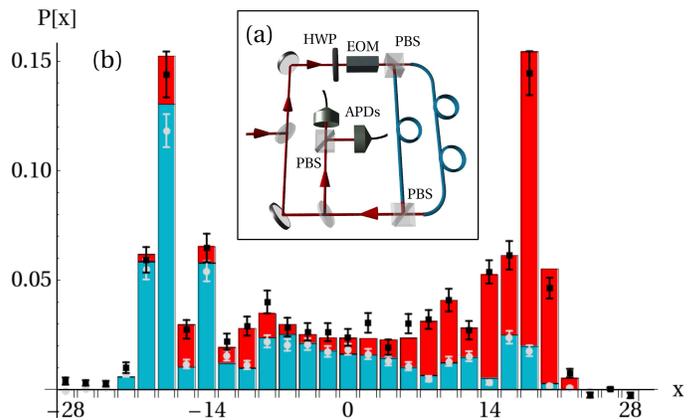}
\caption{(a) Schematic setup. The coin state is manipulated via half-wave plate (HWP) and electro-optic modulator (EOM). The two polarizing beam splitters (PBS) allow to implement the step operation. APDs: Avalanche photodiodes; (b) Probability distribution after 28 steps of a symmetric Hadamard walk with initial circular polarization. Stacked bars: Adapted theory splitted into the two coin states $\ket{V}$ (blue, bottom) and $\ket{H}$ (red, top). Gray dots show experimental data for vertical polarization, black dots the sum of both polarizations. Error bars correspond to statistical errors.} 
\label{setup}
\end{figure}

The functional principle of our experimental setup is sketched in Fig. \ref{setup}(a) and is discussed in detail in \cite{Schreiber}. We generate the input photons with a pulsed diode laser with a central wavelength of 805 nm, a pulse width of 88 ps and a repetition rate of 110 kHz. The initial polarization state of the photons is prepared with retardation plates. Each coin operation consists of a polarization rotation, which is realized with a half-wave plate (HWP), and a subsequent phase shift implemented by a fast switching electro-optic modulator (EOM), as described in Eq.(\ref{coin}). The properties of the EOM impose that $\phi_\mathrm{V}(x) /\phi_\mathrm{H}(x)\approx 3.5$. 
The step operation is realized in the time domain via two polarizing beam splitters (PBS) and a fiber delay line, in which horizontally polarized light follows a longer path (Fig. \ref{setup}(a)). The resulting temporal difference of 5.9 ns between both polarization components corresponds to a step in the spatial domain of $x\pm1$. After a full evolution the photon wave packet is distributed over several discrete spatial positions or, equivalently, over respective time windows. For detection the photon gets coupled out of the loop by a beam splitter with a probability of $12\%$ per step. We employ two avalanche photodiodes (APD) to measure the photon's time and polarization properties, which gives information about the number of steps, the specific position of the photon, as well as its coin state. The probability that a photon undergoes a full roundtrip without getting lost or detected is given by $\eta_{setup}= 0.55$ (0.22) without (with) the EOM and the detection efficiency is $\eta_{det}=0.06$ per step. To determine the statistical distribution of one specific step we detected more than $10^4$ events in an overall measurement time of maximally 1 h, limited by the setup stability. This guaranteed an absolute statistical error of the assessed probability at each position of less than $0.01$. An average photon number per pulse at the detected step of less than $\langle n \rangle < 0.003$ ensured a negligible probability of multi-photon events $P(n>1)/P(n=1)< 0.02$.

(i) \textit{Homogeneous lattice.---} In the first of our four quantum walk scenarios we investigate a homogeneous environment, thus testing the intrinsic coherence properties of the setup. The spatial distribution after 28 steps can be seen in Fig. \ref{setup}(b). We used the initial state  $\ket{\psi_0}=\ket{0}\otimes\frac{1}{\sqrt{2}}(\ket{H}+i\ket{V})$ and the Hadamard coin ($\theta=\pi/8$) at each position. The final state clearly shows the characteristic shape of a fully coherent quantum walk: the two pronounced side peaks and the low probability around the initial position. Moreover, the polarization analysis confirms the expected dependence of the particle's final position on its coin state. An adapted theory including only small imperfections of the coin parameter $\theta$, the initial coin state and differential losses between the two polarizations fully explains the final spatial and polarization distribution. The quality of the result can be quantified by the distance $d(P_m,P_{th})= \frac{1}{2}\sum_{x}|P_m(x)-P_{th}(x)|$ between the measured $P_m$ and the theoretical $P_{th}$ probability distributions. It ranges between 0 for identical distributions and 1 for a complete mismatch. The distance of the measured walk to the adapted quantum theory is $d(P_m,P_{qw}) = 0.052 \pm 0.015$. For comparison we calculated the distance to the fully decoherent (classical) scenario, obtaining $ d(P_m,P_{cl}) = 0.661\pm 0.015$. Hence, our result confirms an almost decoherence free evolution after 28 steps.

(ii) \textit{Static disorder.}--- We implemented the evolution of a particle in an environment with static disorder using a quantum walk with variable coin operation. To create a static disorder a coin operation is required, which is position and not step dependent. In our system this is realized by a controlled phase shift $\phi_\mathrm{H/V}(x)$, such that the photon acquires the same phase any instance it appears at position $x$. To generate a random static phase pattern we applied a periodic noise signal to the EOM. The periodicity of the signal was carefully adjusted to ensure that the applied phase shift operation is strictly position dependent. Using different phase patterns at subsequent runs allows to average over various disorders, as considered in the model of Anderson. The strength of disorder is determined by the maximal applied phase shift $\Phi_{max}$, which defines the uniform interval $\phi_\mathrm{V}(x)\in[-\Phi_{max},\Phi_{max}]$, from which the phases are chosen.
\begin{figure}[ht]
\centering 
\includegraphics[width=0.5\textwidth]{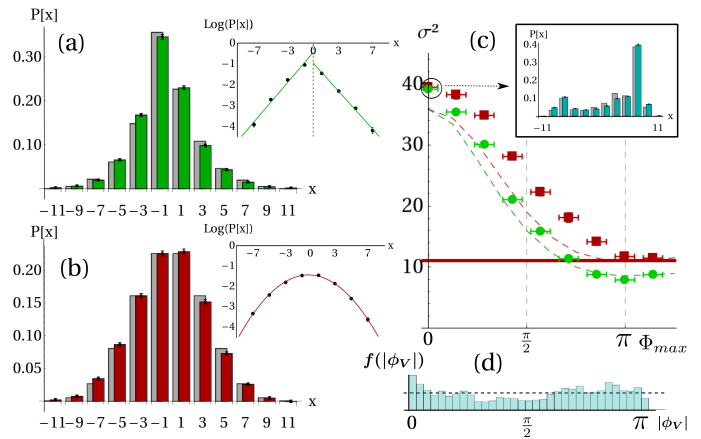}
\caption{Measured probability distribution (front) and respective theory (back, gray bars) of 11 steps of a quantum walk ($\theta=\pi/8$) with static disorder (a), dynamic disorder (b) and in a decoherence free environment (Inset (c)). The insets in (a) and (b) show the measured distribution in semilog scale with linear (a) and parabolic fit (b). (c) Transition of the variance from ballistic quantum walk to diffusive/ localized evolution due to dynamic (red squares) and static (green dots) disorder with increasing disorder strength $\Phi_{max}$; dashed lines: theory without adaption for experimental imperfections. The red solid line marks the variance of a classical random walk. (Vertical error is smaller than the dotsize). (d) Relative frequency $f(|\phi_\mathrm{V}|)$ of the applied phases $\phi_\mathrm{V}$ for the signal with interval $\Phi_{max}= (1.02 \pm 0.05) \pi$. The dashed line indicates the uniform distribution.} 
\label{Anderson}
\end{figure} 
The probability distribution after eleven steps is shown in Fig. \ref{Anderson}(a). We used the initial state  $\ket{\psi_0}=\ket{0}\otimes\ket{H}$, $\theta=\pi/8$ and a high disorder strength ($\Phi_{max} = (1.14\pm0.05) \pi$). In contrast to the decoherence free quantum walk ($\Phi_{max} = 0$, inset of Fig. \ref{Anderson}(c)), in the disordered scenario the expansion of the wave packet is highly suppressed. We observe a strictly enhanced arrival probability around the initial position, which also displays the predicted exponential decay. This striking signature of Anderson localization is emphasized by linear fits in the semilog scaled plot (inset of Fig. \ref{Anderson}(a)). Our results are in agreement with a theoretical  model determined by a Monte Carlo simulation of $10^4$ different phase patterns compatible with our experiment. Compared to (i), the number of steps is reduced due to the additional losses introduced by the EOM.

(iii) \textit{Fast fluctuations.}--- To generate a system with dynamic disorder we detuned the temporal length of the noise signal, thus eliminating position dependent phase correlations. Decoherence appears as a consequence of the dynamically varying phase suffered by the quantum particle during the evolution. As a result, the photon undergoes a classical random walk, revealing a binomial probability distribution (Fig. \ref{Anderson}(b)). In contrast to the previous case, the spatial profile of the wave packet shows a parabolic shape in the semilog scale (inset, Fig. \ref{Anderson}(b)). 

A stepwise increase of the disorder strength $\Phi_{max}$ nicely demonstrates the controlled transition of the system from the ballistic evolution (decoherence free quantum walk) towards the diffusive evolution/localization in a scenario with dynamic/static disorder (Fig. \ref{Anderson}(c)). For this purpose we characterize the resulting expansion profile by its variance $\sigma^2$. Without decoherence ($\Phi_{max} = 0$) the ballistically spreading wave packet shows a large expansion induced by quantum interference after eleven steps. The slightly higher variance in comparison to the theory can be explained by small polarization rotations in the EOM. In a system with dynamic disorder, decoherence reduces the expansion of the wave packet to the level of a diffusive classical particle. In contrast, static disorder leads to a stagnation of the spread and hence an even smaller variance. Our results clearly demonstrate how the amount and kind of disorder influence the expansion of the particle's wave packet.

The agreement between theory and measurement in the completely dephased scenario (Fig. \ref{Anderson}(b)) confirms a sufficient randomness of the applied noise signal. Furthermore, an independent interferometric measurement revealed the relative frequency of the used phases $f(|\phi_\mathrm{V}|)$, as can be seen in Fig. \ref{Anderson}(d) with $\Phi_{max}= (1.02 \pm 0.05) \pi$.  However, small imperfections of the EOM lead to a distribution slightly off uniformity, which also influences the measured variances shown in Fig. \ref{Anderson}(c).

\begin{figure}
\centering 
\includegraphics[width=0.5\textwidth]{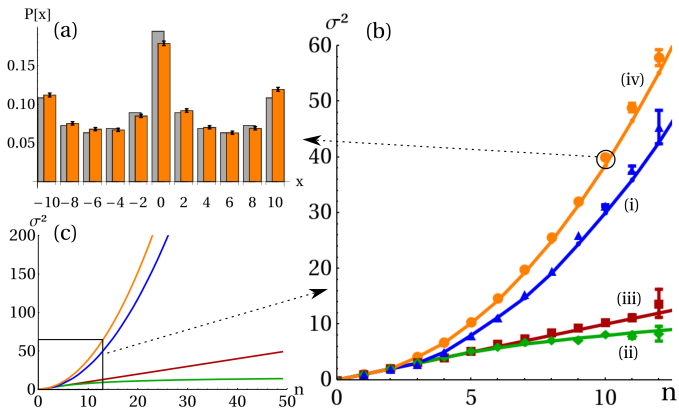}
\caption{(a) Averaged probability distribution in a slow decoherence scenario with different coin angles $\theta \in [0, \pi/4]$: Measurement (orange, front) and theory (gray, back); (b)+(c) Trend of the variance with the number of steps: Measurement up to 12 steps (b) and simulation up to 50 steps (c). While photons undergoing the decoherence free quantum walk (blue triangles) and the evolution with slow dynamic disorder (orange dots) show a ballistic behavior, in a classical random walk (red squares) they move diffusively, and, finally, in the case of static disorder (green diamonds) they stagnate.} 
\label{deviation}
\end{figure}

(iv) \textit{Slow fluctuations.}--- As the fourth scenario we simulated fluctuations in a homogeneous system, but with parameters that change in a time scale much larger than the full duration of a single quantum walk. Although the individual evolution is not affected under these circumstances, an ensemble measurement of subsequent walks results in an average over coherent evolutions in different types of lattices. For this purpose we changed the parameter $\theta \in [0, \pi/4]$ in steps of $\pi/18$ for a quantum walk with initial state $\ket{\psi_0}=\ket{0}\otimes\frac{1}{\sqrt{2}}(\ket{H}+i\ket{V})$. An average over the full range $\theta \in [0, \pi/4]$ exhibits a nearly uniform spatial distribution of the wave packet with an enhanced probability to arrive at its initial position $x=0$ after 10 steps (Fig. \ref{deviation}(a)). Especially the high chance to reach the outermost positions $x=\pm10$ differs significantly from all previous scenarios. This increases the variance of the distribution ($\sigma^2_{\mathrm{(iv)}}= 40.00 \pm 0.42$) to a level, which is even higher than in the decoherence free quantum walk with the Hadamard coin ($\sigma^2_{\mathrm{(i)}}= 31.27 \pm 0.19$). The result demonstrates that special kinds of decoherences can even speed up the expansion of wave packets in homogeneous lattices.

Finally, the geometry of the setup allows to observe easily the wave packet's evolution step by step in all four scenarios (Fig. \ref{deviation}(b)). For cases (i) and (iv) we observe a ballistic spread, with an even faster expansion in a system with slow fluctuations. The evolution with fast dynamic disorder (iii) is clearly diffusive. Lastly, under the condition of static disorder (ii) the variance saturates after few steps and the dynamics is dominated by the effect of Anderson localization. For comparison, we show in Fig. \ref{deviation}(c) a theoretical plot for the evolution of the variance over fifty steps. The parameters used in simulation and experiment are equivalent to the experimental settings used for Figs. \ref{setup}(b), \ref{Anderson}(a-b) and \ref{deviation}(a).

In conclusion, we presented how disorder and fluctuations in a periodic lattice can influence the evolution of a traversing particle. We observed a fast ballistic spread for slowly changing lattice parameters, a diffusive spread in the case of dynamical disorder and Anderson localization for lattices with static disorder. Furthermore, we showed the controlled transition between the different regimes. The high flexibility and control allows not only the study of further decoherence phenomena in quantum walks but also to simulate specific physical scenarios of interest for the solid state and biophysics community. Moreover, the possibility to manipulate quantum walks with time dependent coin operations is a fundamental step towards the realization of quantum walk-based protocols. 

We thank P.P. Rohde for helpful discussions.
We acknowledge financial support from the German
Israel Foundation (Project 970/2007). K.N.C. and I.J.
acknowledge support from the Alexander von
Humboldt Foundation; V.P., A.G. and I.J. from MSMT
LC06002 and MSM 6840770039.

\bibliographystyle{unsrt}

\begin{thebibliography}{10}



\bibitem{Aharonov}
Y.~Aharonov, L.~Davidovich, and N.~Zagury.
Phys. Rev. A \textbf{48}, 1687 (1993).

\bibitem{Childs}
A.~M. Childs,
Phys.Rev. Lett. \textbf{102}, 180501 (2009).


\bibitem{Mohseni}
M. Mohseni \textit{et al.},
J. Chem. Phys. \textbf{129}, 174106 (2008).

\bibitem{Aspuru-Guzik}
P. Rebentrost \textit{et al.},
New J. Phys. \textbf{11}, 033003 (2009).

\bibitem{Whaley}
S. Hoyer, M. Sarovar, and K. B. Whaley,
 New J. Phys. \textbf{12}, 065041 (2010).

\bibitem{Meschede}
M. Karski \textit{et al.},
Science \textbf{325}, 174 (2009).

\bibitem{Schaetz}
H. Schmitz \textit {et al.},
Phys. Rev. Lett. \textbf{103}, 090504 (2009).

\bibitem{Blatt}
F.~Z\"{a}hringer \textit{et al.},
Phys. Rev. Lett. \textbf{104}, 100503 (2010).

\bibitem{NMR}
J. Du {\it et al.}, Phys. Rev. A \textbf{67}, 042316 (2003). 

\bibitem{NMR2}
C. A. Ryan {\it et al.},  Phys. Rev. A \textbf{72}, 062317 (2005).


\bibitem{Silberberg2}
H. B. Perets {\it et al.},
Phys. Rev. Lett. \textbf{100}, 170506 (2008).

\bibitem{White}
M.~A. Broome \textit{et al.},
Phys. Rev. Lett. \textbf{104}, 153602 (2010).

\bibitem{Schreiber}
A. Schreiber \textit{et al.},
Phys. Rev Lett. \textbf{104}, 050502 (2010).

\bibitem{OBrien}
A. Peruzzo {\it et al.},
Science \textbf{329}, 1500 (2010).

\bibitem{SKW}
N. Shenvi, J. Kempe and K. B. Whaley,
Phys. Rev. A \textbf{67}, 052307  (2003).

\bibitem{Ambainis}
A. Ambainis,
SIAM Journal on Computing, \textbf{37}, 210-239 (2007).

\bibitem{Anderson}
P.~W. Anderson,
Phys. Rev. \textbf{109}, 1492 (1958).

\bibitem{Righini}
D.~S. Wiersma \textit{et al.},
Nature \textbf{390}, 671 (1997).

\bibitem{Silberberg}
Y. Lahini \textit{et al.},
Phys. Rev. Lett. \textbf{100}, 013906 (2008).

\bibitem{Schwartz}
T. Schwartz \textit{et al.},
Nature \textbf{446}, 52 (2007).

\bibitem{Roati}
G. Roati \textit{et al.},
Nature \textbf{453}, 895 (2008).

\bibitem{Aspect}
J. Billy \textit{et al.},
Nature \textbf{453}, 891 (2008).

\bibitem{Toermae}
P.~T\"{o}rm\"{a}, I.~Jex, and W.~P. Schleich,
Phys. Rev. A \textbf{65}, 052110 (2002).

\bibitem{Winter}
J.~P. Keating \textit{et al.},
Phys. Rev. A \textbf{76}, 012315 (2007).

\bibitem{Evangelou}
Yue Yin, D. E. Katsanos, and S. N. Evangelou,
Phys. Rev. A \textbf{77}, 022302 (2008).

\bibitem{Ambainis 2}
T. A. Brun, H.~A.Carteret, and A. Ambainis,
Phys. Rev. Lett. \textbf{91}, 130602 (2003).

\bibitem{Kendon}
V. Kendon,
\newblock { Math. Struct. Comp. Sci.} \textbf{17}, 1169 (2007).






\end{thebibliography}

\end{document}